\documentclass[aps,prl,preprint,showpacs,floatfix]{revtex4}
\usepackage{graphicx}

\begin{document}

\title{ Photoinduced Suppression of Superconductivity in Bi$_2$Sr$_2$Ca$_2$Cu$_3$O$_{10+\delta}$}

\author{ Subhasis Ghosh and Niladri Sarkar}

\affiliation{School of Physical Sciences, Jawaharlal Nehru University, New Delhi 110067}

\begin{abstract}

Superconductivity in  high temperature superconductors is achieved by introducing charge carriers  into cuprate  insulators containing CuO$_2$ planes.  Perturbation in  these CuO$_2$ planes   suppresses superconductivity by strongly affecting the electronic environment.   Here we have use photoinduced electronic modification of Cu ion in CuO$_2$ plane to investigate the effect of local perturbation in Bi$_2$Sr$_2$Ca$_2$Cu$_3$O$_{10+\delta}$. This method has been used to suppress  superconducting transition temperature T$_c$. We show that our results on photoinduced suppression of superconductivity are consistent with  a scenario based on pinning of fluctuating stripes  in  Bi$_2$Sr$_2$Ca$_2$Cu$_3$O$_{10+\delta}$.

\end{abstract}


\maketitle

Superconductivity in high temperature superconductors(HTSC) is generally believed to originate from the strongly interacting carriers(holes or electrons) in  CuO$_2$ planes, but, the exact mechanism of high temperature superconductivity   has been debated intensely since it was discovered.  One of the most fascinating properties of HTSC is the sensitivity of the critical temperature, T$_c$, at which metal becomes superconducting, to doping and non-magnetic(Zn, Li, Al) impurities, which are substituted with Cu atoms in the CuO$_2$ planes.  The effect of local perturbation in CuO$_2$ planes, in particular, the suppression of superconductivity by nonmagnetic impurities in CuO$_2$ planes is being studied experimentally\cite{ha91,avm94,yf96,bn96,jb99,shp00} and theoretically\cite{nn97,ahc98,ap01,cms01,tam02,pp04}.  The suppression of T$_c$ has been explained by different scenarios, such as modifications of local magnetic environment\cite{avm94}, pair breaking by strong potential scattering in the presence of {\sl d}-wave order parameter\cite{rew93},  destruction  of superfluid density  in ``swiss chesse''-like  void created by impurity\cite{bn96}, impurity induced local variation of the superconducting gap\cite{mf97},   Kondo impurity scattering\cite{nn97,ap01},   pinning of stripe by planar impurity\cite{ahc98,cms01} and enhancement of antiferromagnetic spin correlation\cite{tam02}.

Nevertheless, the understanding of the impact of such an impurity on HTSC in normal and superconducting state, in particular,  the  microscopic origin  of suppression of superconductivity is far from satisfactory.    In this letter, we present results on  the photoinduced T$_c$ suppression in  Bi$_2$Sr$_2$Ca$_2$Cu$_3$O$_{10+\delta}$(BSCCO-2223). The experiments were performed on high quality superconducting platelets of BSCCO-2223. The maximum grain misalignment angle in the powder-in-tube(PIT) processed tape, from which it was extracted, was around  8$^o$. This enabled the supercurrent flow to be primarily lossless across the grain boundaries of the tape\cite{dcl01}.  Typical thickness of the sample was 100$\mu$m.  The samples have been chosen carefully with regard to sharp superconducting transitions of width$\sim 1K$. The slightly underdoped($T_c\sim100K$) samples are chosen for  minimizing the effect of pseudogap  and other competing orders and  to have linear temperature dependence of resistivity $\rho(T)$.   There are several disadvantages with Zn-doped and irradiated samples. In both cases, T$_c$-suppression is also affected by uncontrolled and unwanted defects. Resistivity was measured during heating and cooling. The   heating and cooling rates($\sim$0.1K/min)    were adjusted  very slowly to avoid any thermal  effects\cite{lrt71}.  {\sl The main advantage in this case, in contrast to impurity doped or irradiated samples,   T$_c$-suppression can be  achieved reversibly}. We have explained the  suppression of superconductivity  with  a scenario based on pinning of fluctuating stripes by photoinduced Cu ion in $3d^{10}$ state(Cu$+$).    

The evolution of T$_c$ as a function of intensity of light  is shown in Fig.1,  displaying  the parallel shift of resistivity $\rho(T)$ curves, which show linear temperature dependence following the relation, $\rho(T) = \rho_0 + \alpha T$.   There is no signature of pseudogap related downward deviation of $\rho(T)$\cite{tw97}. In order to estimate the effects of photoexcitation on $\rho(T)$, we looked at the slope $\alpha = \frac{d\rho}{dT}$, which depends on the hole concentration $p$.  It is clear from Fig.1 that the slope $\alpha$ remains almost same for all resistivity curves.  Hence, the photoexcitation does not change the effective hole concentration, it only modifies the residual resistivity $\rho_0$, which is a measure of the scattering rate of charge carriers by impurities. This reduction of T$_c$ can be interpreted as the suppression of superfluid density $\sigma_s$, as $T_c \propto \sigma_s$\cite{bn96,yju89}.   Fig.2 shows how the suppression of T$_c$, i.e. $\Delta T_c$ depends on wavelength of light. It should be mentioned that the parent compounds are not strictly Mott insulator  {\sl per se} and are more accurately described as charge transfer(CT) insulator.  In this case O $2p$ band lies above Cu $3d$ LHB and minimum band gap is not the correlation energy $U$, but the gap between UHB and O $2p$ band, known as CT gap, which is around 1.5-2eV in cuprate HTSCs\cite{yo91}. So, one would expect the maximum of suppression of T$_c$ should be at CT gap in HTSC, instead, we have observed suppression of T$_c$ starts at 2eV, which is the charge transfer gap in  BSCCO-2223\cite{yo91} and  maximum of $\Delta T_c$  is at 2.6eV and not at CT gap.     We attribute this transition  due to  Cu-3d$^9$(Cu$^{2+}$) to Cu-3d$^{10}$(Cu$^{+}$), known as $d^9-d^{10}L$ transition\cite{mjh94},  photoexcitation  makes some  Cu ions in CuO$_2$ planes  as spinless  non-magnetic potential perturbation as Zn$^{2+}$ does in CuO$_2$ plane in case Zn-doped cuprates.    So, what is the origin of this spectral dependence?  The superconductivity induced  changes in optical spectra  at energies more than CT gap has been established\cite{mr01}  in Bi$_2$Sr$_2$CaCu$_2$O$_{8+\delta}$(BSCCO-2212)    using spectroscopic ellipsometry.  Corroborating with neutron scattering experiments, this spectral feature   has been connected with magnetic degrees of freedom. This connection is also supported by inelastic light scattering related to two-magnon resonance energies at higher than CT gap  in BSCCO-2212. In addition to this,  similar resonance at 2.7eV have been observed\cite{gb97,mr99} in Raman spectra  for Cooper pair breaking in BSCCO-2212  over a wide range of doping. All these suggest that  there have to be electronic states strongly coupled to magnetic degrees of freedom and  superconductivity is  induced by the interaction with the high energy magnetic excitations.

To compare our result with different  mechanisms for destruction of superconductivity, we have studied how $T_c$ changes with intensity of light or the concentration of photoexcited singly positive Cu ions(n$^{Cu+}$). The best estimate of  n$^{Cu+}$  can be obtained from photoinduced change in residual resistivity  $\Delta\rho^{o}_{ab}$ = $\rho_0^{ph}$ -  $\rho_0^d$, where $\rho_0^d$ and $\rho_0^{ph}$  are the  residual resistivities in dark and in the presence of photoexcitation, respectively.  Fig.3 shows the linear variation of $T_c$ with $\Delta\rho^{o}_{ab}$. Generally it has been always attempted  to explain the destruction of superconductivity by Abrikosov-Gorkov(AG) relation\cite{aaa61}, which is  weak coupling formalism based on Bardeen-Cooper-Schieffer(BCS) theory for   elastic scattering due to impurities embedded in a BCS host.   This relation is given as $\ln \left ( \frac{T_{cd}}{T_c}\right ) = \psi \left (\beta + \frac{1}{2}\right) - \psi \left (\frac{1}{2}\right)$,   where $\psi(x)$ is digamma function and $\beta = \frac{\hbar}{2\pi k_B T_c \tau}$, $\tau$ is the scattering rate and  T$_{cd}$ is the superconducting transition temperature in dark. Now, $\tau$ can be expressed in terms $\rho_{0}^d$ as, $\frac{1}{\tau}=\frac{\rho_{0}^d\omega^{2}_{p}}{4\pi}$,  
where $\omega_{p}$ is the plasmon frequency.   Fig.3 shows the fitting for suppression of T$_c$ with AG relation with different values of $\omega_{p}$ in BSSCO\cite{ib90}.  It is clear that experimental data cannot be  explained by AG relation.     Any attempt to reproduce linear variation of T$_c$ with residual resistivity  by BCS based AG relation or it's variant(such with d+s order parameter\cite{lao04} or spatial variation of order parameter\cite{mf97})  resulted   unacceptable and unphysical values of different parameters such as plasmon frequency and transport coupling constants. This is not surprising because  these  methods employed to determine the suppression of $T_c$ cannot be satisfactorily described  within the framework of a  BCS-like mean field theory. It is clear from Fig.3 that the suppression of $T_c$ is linear with residual resistivity  following the relation,

\begin{equation}
\frac{T_c}{T_{cd}} = 1 - \frac{\rho_0^{ph}}{\rho_0^c}
\end{equation}

\noindent  In this case $T_c$ is a function of light intensity and can be scaled with residual resistivity,  as $\rho_0$ and $\alpha(=\frac{d\rho}{dT})$ dont' change with light intensity. $\rho_0^c$ is the critical residual resistivity at which the superconductivity is completely destroyed.   By comparing the best fit to the data, we determine $\rho_0^c$=390$\mu\Omega$cm, which should be the  critical residual resistivity when superconductivity is completely destroyed.

Now we can estimate the number of photoinduced spinless Cu ions(3d$^{10}$),  n$^{Cu+}$,  from the photoinduced variation of the 2D residual resistivity $\rho^{2D}_{o}$, which is given by\cite{nn97}, $\rho^{2D}_0=\frac{4\hbar}{e^{2}}\frac{n_{imp}}{\delta}$, 
where $\delta$ is hole concentration and  $n_{imp}$ is the concentration of impurity.  As already discussed in the previous section, photoexcitation transforms some Cu ions in CuO$_2$ planes from Cu-3d$^9$(Cu$^{2+}$) to spinless Cu-3d$^{10}$(Cu$^{+}$). These Cu ions in CuO$_2$ planes act as spinless  non-magnetic potential perturbation as Zn$^{2+}$ does in CuO$_2$ plane in case Zn-doped cuprates. 
Therefore, the term $n_{imp}$  becomes $n_{imp}$+n$^{Cu+}$ in the presence of photoexcitation and  the value of $\rho^{2D}_{o}$ at any light intensity can be given by, $\rho^{2D}_{o}=\frac{4\hbar}{e^{2}}\frac{(n_{imp}+n^{Cu+})}{\delta}$. From Fig.1 we see that there is a parallel shift of resistivity $\rho(T)$ curves with light intensity. As we have mentioned, photoexcitation does not change the effective hole concentration, so the value of $\delta$ remains constant and the only parameter that varies with light intensity is  n$^{Cu+}$, which has been determined from  $\rho_0^{2D}$(Fig.3). 

The  problem of HTSC has been addressed as behavior of doped holes(electrons) in an antiferromagnetic background and shown that interplay between superconductivity and antiferromagnetism can give rise to inhomogeneous spin and charge ordering\cite{sak03,ewc03}, known as ``stripes''. Essentially the underlying physics behind stripe formation is the expulsion of holes from the regions of local moments.  There is strong evidence for metallic stripe in La-Sr-Cu-O(LSCO) family of HTSC\cite{jmt95,mf02}. There are some reports on the observation of  stripes in BSCCO-2212\cite{am02,jeh02}. It has been argued strongly that dynamic stripes  underly the superconducting state and shown that suppression of $T_c$ can be explained by pinning the fluctuating stripes\cite{sak03,jmt95,mf02,ahc98,cms01}.  
In this case  Cu$^+$ ions  pin the  stripes and slow their fluctuation by reducing the transverse kinetic energy. It has been shown by neutron scattering experiments on La$_2$CuO$_{4+y}$ that distance between stripes is not affected by planar impurity with closed d-shell\cite{bow97}. Hence the stripe pinning by Cu$^+$ ion must be local within a sphere of influence around the Cu$^+$, resulting  a  reduction of $T_c$. Fig.4  schematically shows the effect of  Cu$^+$ ion on the stripe grid.   In the presence of Cu$^+$ ions,   fluctuations are suppressed close to the  Cu$^+$ site. Following Refs.8 and 10,  we assume that the Cu$^+$ions are located half the distance between superconducting stripes and suppress stripe fluctuations within a circle of radius R around their position, as shown in Fig.4.  This distance R is set by various energy scales in the problem\cite{ahc98,cms01} including the string tension of the stripe and therefore should scale with the interstripe distance {\sl l}. From Fig.4 we see that the linear region of the stripes that is affected is $\sqrt{R^{2}-{\sl l}^{2}/4}$, where  $R=\gamma{\sl l}/2$. Here $\gamma$($\gamma$$>$1) is a doping independent phenomenological parameter which depends on the disorder and energetic details\cite{ahc98,cms01}. Now,
the stripe length which is pinned per $Cu^{+}$ ion is $4\sqrt{R^{2}-{\sl l}^{2}/4}$=$2{\sl l}\sqrt{\gamma^{2}-1}$. Pinning of fluctuating stripes  leads to suppression of superfluid density  and if 
 there are $n_{Cu^{+}}$ photoinduced ions per unit volume, and if we assume that all these ions take part in pinning the stripes, then the  planer density of suppressed superfluid density $\Delta \sigma_s$ is given  $\Delta \sigma_s = \frac{2n^{Cu^{+}}{\sl l}^{2}\sqrt{\gamma^{2}-1}}{a^{2}}\sigma_s$, where $a$ is lattice constant. Finally, the net superfluid density in the presence of photoexcitation  is given by $\sigma_s^{ph}=\sigma_s^d - \Delta \sigma_{s}$.  Now, using  Uemura's relation\cite{yju89} and Eqn.1,  we have

\begin{equation}
\frac{\sigma_s^{ph}}{\sigma_s^d}=1-\frac{2n^{Cu^{+}}{\sl l}^{2}\sqrt{\gamma^{2}-1}}{a^{2}} =1-\frac{n^{Cu^{+}}}{n_{C}^{{Cu^{+}}}}=1 - \frac{\rho_0^{ph}}{\rho_0^c}=\frac{T_{c}}{T_{cd}}
\end{equation}

\noindent  where $n_{C}^{{Cu^{+}}}$ is the critical concentration of $Cu^{+}$ ions when superfluid density vanishes and given by  $n_{C}^{{Cu^{+}}}=\frac{1}{2\sqrt{\gamma^{2}-1}}\left(\frac{a}{{\sl l}}\right)^{2}$. 
 Fig.3  shows the variation of the $T_{c}/T_{cd}$ with  n$^{Cu+}$. The suppression of $T_{c}$ is linear with the n$^{Cu+}$, as predicted  by Eqn.2 and  the value of $n_{C}^{{Cu^{+}}}$ has been found to be  0.0373. Using the  relation  $\frac{1}{2\sqrt{\gamma^{2}-1}}\left(\frac{a}{{\sl l}}\right)^{2}=0.0373$, one obtains $l=4.7a$  and $R=14.84\AA$, where $a=5.4\AA$ and $\gamma = 1.17$. 
Similar value of R is reported\cite{shp00} in case of BSCCO-2212, where scanning tunnelling microscopy(STM) was used to investigate the effects of individual zinc impurity atoms. It has been observed that there is a strong suppression of superconductivity within $\sim$15$\AA$ of Zn sites in BSCCO-2212.

Finally,  we would like to mention that it is clear in Fig.1, while  $\rho$ keeps decreasing  over the whole temperature range above $T_c$, there appears a feature below $T^* \sim$ 260K in resistivity curve $\rho(T)$. This feature becomes pronounced in resistivity curves $\rho(T)$ taken under photoexcitation. In this case resistivity  deviates upward from the linearity as $T$ decreases and finally $\rho(T)$ resumes linear-$T$ descent. This is different from well known downturn deviation observed in underdoped cuprates due to pseudogap\cite{tw97}.  This upward deviation has been observed in LSCO\cite{tn99}, BSCCO-2223\cite{wc99} and attributed to high temperature tetragonal(HTT) to the low temperature orthorhombic(LTO) phase transition. This led to the speculation that the upward deviation of $\rho(T)$ is universally observed in cuprates, and is related  to the lattice distortion induced structural phase transition and an  indication of formation of  stripe phase, which has been observed in LSCO and YBCO and only speculated about its' existence in BSCCO.

In conclusion, we have performed reversible experiments on destruction of superconductivity using photoexcitation in slightly underdoped BSSCO-2223. The reported results  not only reveals new physics regarding the origin of suppression of  superconducting in HTSC  but also opens a new avenue for investigating the role of disorder or perturbation in copper oxide plane for  understanding the mechanism of  superconductivity in HTSCs.

\newpage

\noindent {\bf Figure Captions}

\begin{description}

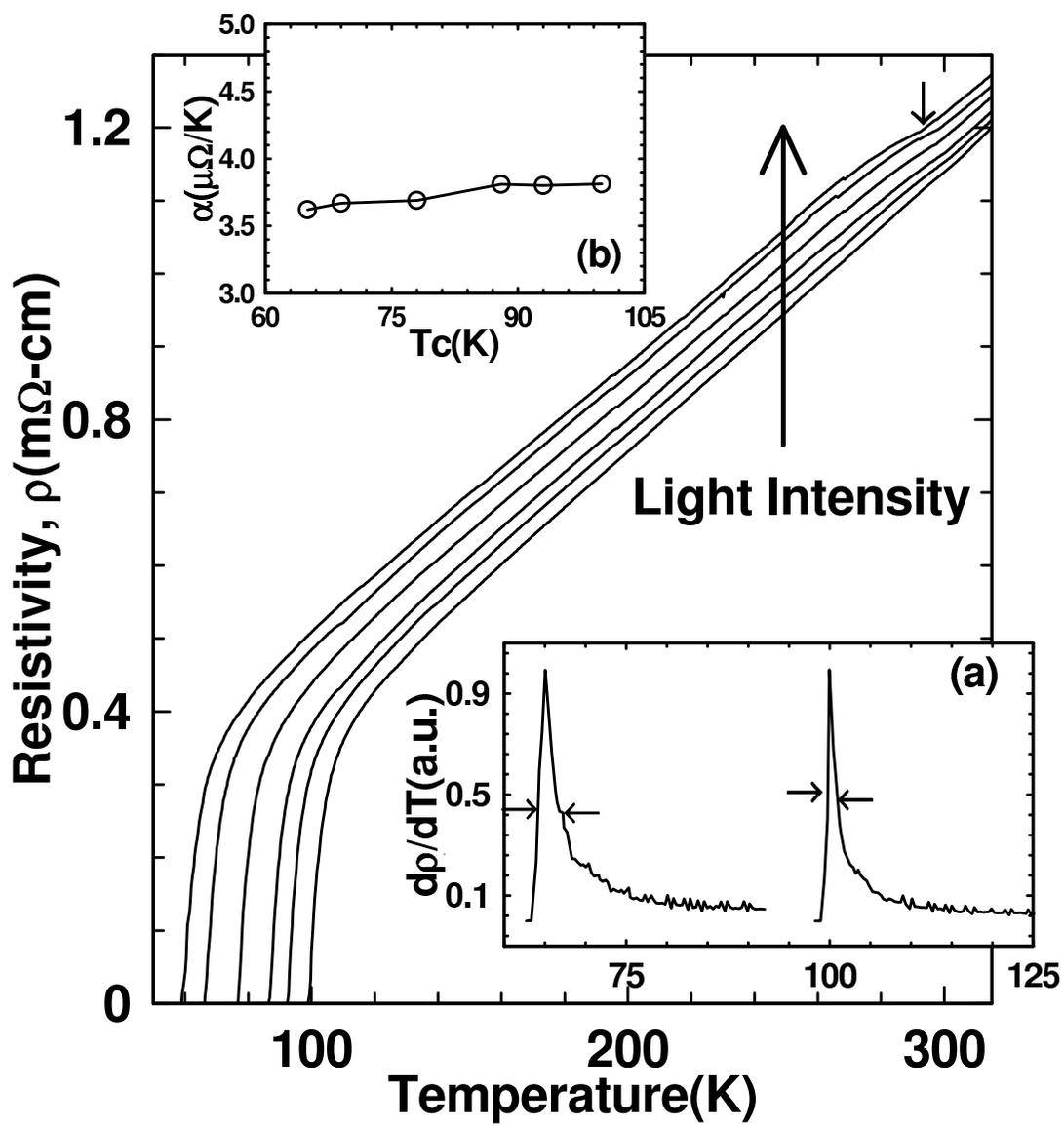
\item{Figure 1.} Temperature dependence of in plane resistivity for slightly underdoped BSSCO-2223  in the presence of photoexcitation with 488nm laser light. T$_c$ shifts from 100K(in dark) to 65K in the presence of maximum photoexcitation. Maximum intensity of laser was 20mW/cm$^2$. The downarrow($\downarrow$) shown on the resistivity curves at 260K indicates the point of upward deviation from linear temperature dependence of $\rho(T)$. (a) $d\rho/dT$ near transition temperature in case of T$_c$=100K and T$_c$=65K. Transition width $\delta T_c$  is 1.2K in case of dark and changes to 1.4K in case of highest photoexcitation. Hence, photoexcitation does not induce disorder or inhomogeneity as $\delta T_C$ is the measure of disorder induced inhomogeneities in the sample.   (b) Slope of $\rho(T)$  in linear region(140K to 260K)is plotted against the T$_c$. Slope almost remain unchanged,  but T$_c$ varies from 100K to 65K. Connecting line is guide for eyes.

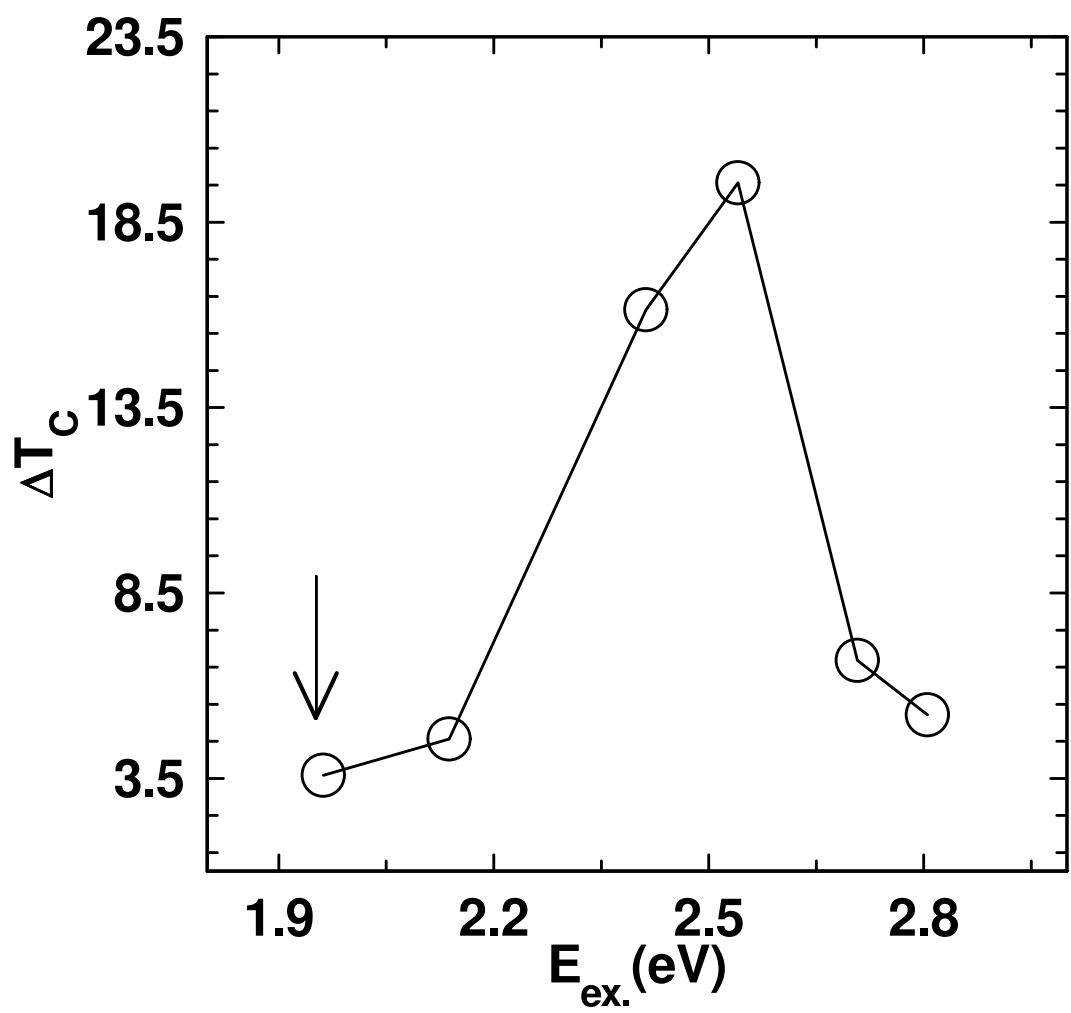
\item{Figure 2.} Suppression of superconducting temperature $\Delta T_c$ is plotted against the energy of photoexcitation. Downarrow($\downarrow$) shows the position of the charge transfer(CT) gap in BSSCO-2223. In this case intensity of light has been fixed to a optimum value(10mW/cm$^2$) and $\Delta T_c$ is measured as the wavelength of light  is changed.

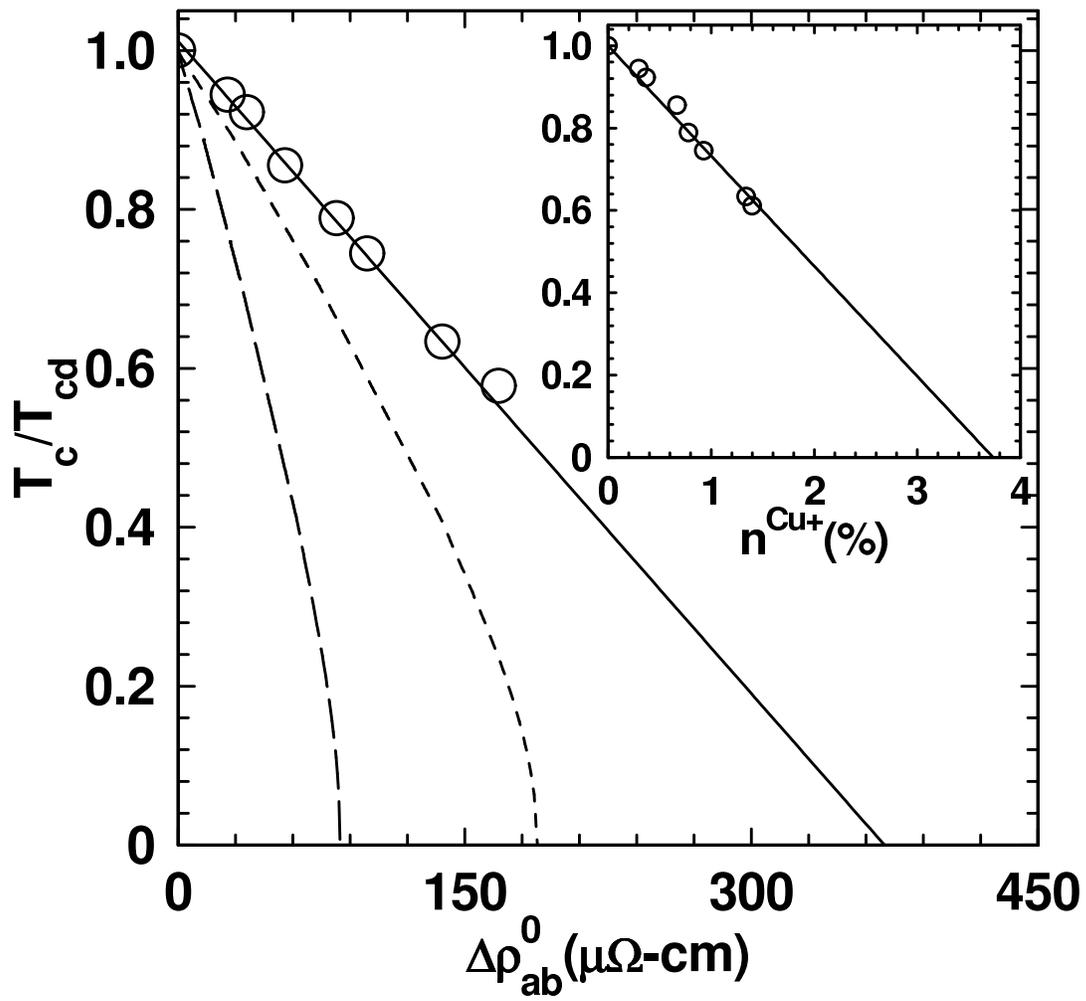
\item{Figure 3.} Reduction of T$_c$ is plotted against the change of residual resistivity. T$_c$ is normalized with T$_{cd}$(T$_c$ in dark). Bigger dashed and smaller dashed  curves represent fitting with AG's relation for two different values of plasmon frequencies of $\omega_p$=1.2eV and $\omega_p$=0.8eV, respectively. Linear fit is according to Eqn.1. Inset shows the normalized T$_c$ at different light intensities vs. the number of photoinduced spinless Cu$^+$(3d$^{10}$) ions determined from the residual resistivity.

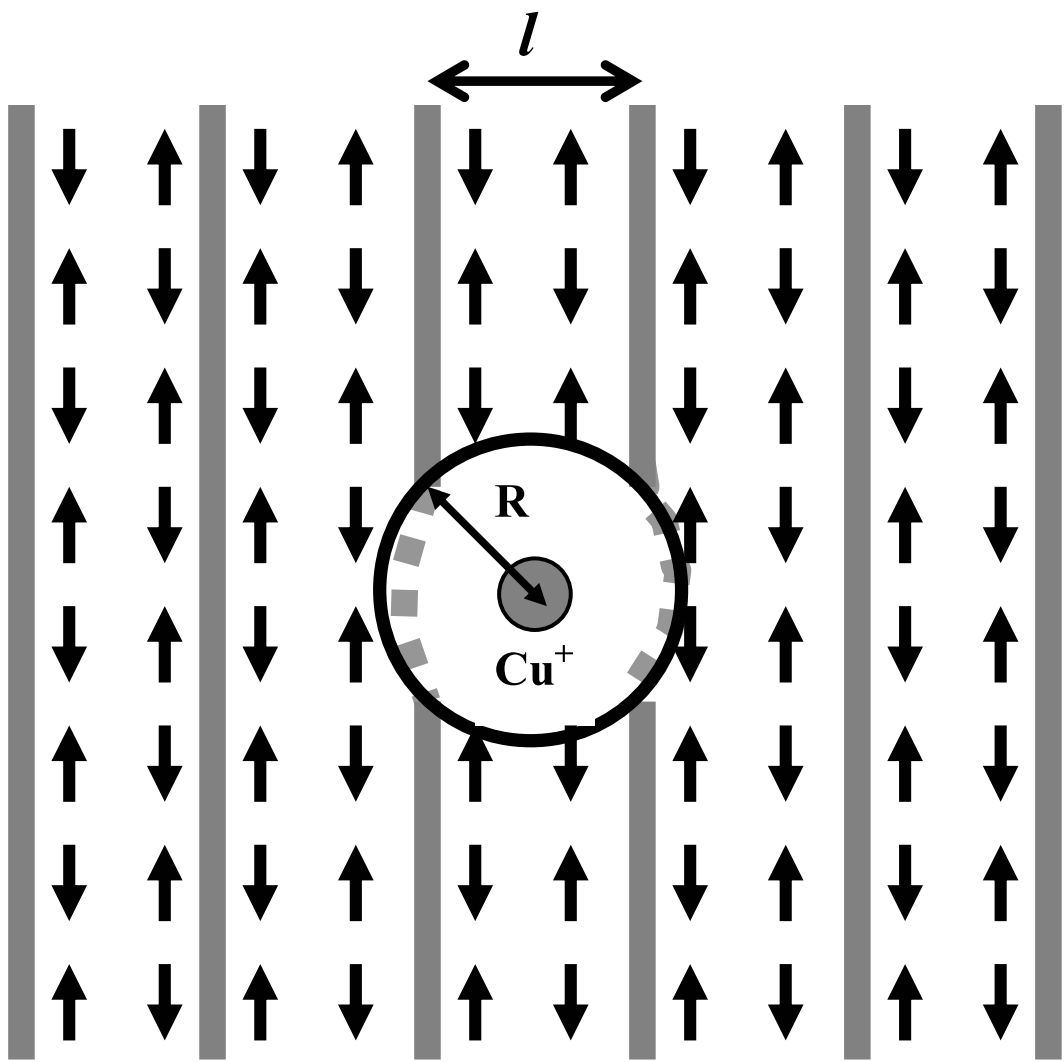
\item{Figure 4.} Schematic diagram of stripes in CuO$_2$ planes. Thick grey lines represent the regions of stripe with finite superfluid density, separated with regions of localized spins. Photoinduced Cu$^+$(3d$^{10}$) pins the stripes within a radius of R, where superfluid density is suppressed.

\end{description}

\end{document}